# Electronic structure and Fermi surface of iron-based superconductors $R_2Fe_3Si_5$ ($R$ = Lu;Y;Sc) from first principles

M. Samsel-Czekała and M.J. Winiarski

*Institute of Low Temperature and Structure Research, Polish Academy of Sciences, P.O. Box 1410, 50-950 Wrocław 2, Poland*

Electronic structures of three superconducting rare-earth iron silicides $(Lu;Y;Sc)_2Fe_3Si_5$ and non-superconducting $Lu_2Ru_3Si_5$, adopting a tetragonal crystal structure ($P4/mnc$), have been calculated employing the full-potential local-orbital method within the density functional theory. The investigations were focused particularly on the band structures and Fermi surfaces, existing in four bands and containing rather three-dimensional electronlike and holelike sheets. They support an idea of unconventional multi-band superconductivity in these ternaries, proposed earlier by other authors for $Lu_2Fe_3Si_5$, based on heat-capacity, resistivity, electromagnetic and muon spin rotation measurements. Finally, a discussion on differences in the electronic structures between the investigated here and other common families of iron-based superconductors is carried out.

## 1. Introduction

In Iron-based superconductors draw wide interest because of recently discovered high-temperature (high-$T_C$) superconducting rare-earth (oxy)pnictides, like $SmFeAsO_{1-x}F_x$ and $Sr_{1-x}Sm_xFeAsF$, reaching the highest transition temperatures, $T_C$, of 55 K due to doping [1]. Their crystal structures are strongly anisotropic, quasi-two-dimensional (quasi-2D), being built from negatively charged PbO-type layers of iron and non-metallic atoms as well as positively charged layers of either alkaline or rare-earth atoms. However, the first known group of superconducting rare-earth iron-based compounds is the studied here family of $R_2Fe_3Si_5$ ($R$ = Lu;Y;Sc) [2,3]. Its members crystallize in the tetragonal structure of the $Sc_2Fe_3Si_5$ type ($P4/mnc$, space group no. 128), containing iron atoms arranged both in squares within planes perpendicular to the $c$ axis and in quasi-1D chains along this axis. The iron planes are lying much closer to one another than those in the (oxy)pnictides or chalcogenides, which yields a more three-dimensional (3D) configuration. These ternary iron silicides exhibit relatively low values of $T_C \le 6.2$ K. Nevertheless, a comparison between their electronic structures and those in high-$T_C$ (oxy)pnictides may be useful in understanding a mechanism of superconductivity (SC) in various groups of iron-based systems.

The majority of the $R_2Fe_3Si_5$ (where $R$ = rare-earth) series order antiferromagnetically with the magnetic moments originating only from the lanthanide $R$ atoms [4-10], except for just investigated in this paper paramagnetic superconductors, namely $Lu_2Fe_3Si_5$ ($T_c$ = 6.25 K), $Y_2Fe_3Si_5$ ($T_c$ = 1.68 K), and $Sc_2Fe_3Si_5$ ($T_c$ = 4.46 K) [2,3,11,12]. Interestingly, a separation between two different antiferromagnetic phases and the superconducting low-temperature phase ($T_c$ = 0.47 K) occurs in $Er_2Fe_3Si_5$ [13]. Finally, $Tm_2Fe_3Si_5$ becomes under pressure a reentrant superconductor, in which the superconducting state is destroyed at the antiferromagnetic transition. This compound is significantly sensitive to applied pressure and any disorder [14-17].
Meanwhile, in $Lu_2Fe_3Si_5$ a rapid depression of $T_C$ by magnetic impurities has being explained by the effect of screening the Fe 3d electrons (diminishing conductivity) by the $f$-electrons [18-20].

Up to now, the two-gap BCS-like superconductivity model has successfully been applied to $(Lu;Y;Sc)_2Fe_3Si_5$, yielding good agreement with the experimental data [3,21-25]. The anisotropy of their superconducting properties, anomalous upper critical fields (in $Lu_2Fe_3Si_5$), and inter-band electron scattering in the case of two weak-coupled distinct gaps opened on the whole Fermi surface (FS) sheets, indicated that they are rather quasi-2D superconductors [26-29]. Furthermore, some recent works [30-32] focused on the effect of doping by non-magnetic impurities and atomic disorder induced by the neutron irradiation, both causing a fast suppression of $T_C$ in $Lu_2Fe_3Si_5$, have questioned the conventional (phononic) mechanism of SC in the $(Lu;Y;Sc)_2Fe_3Si_5$ family, revealing the significance of spin fluctuations in formation of the SC state [32]. The same effect was observed in the

high-$T_C$ (oxy)pnictides after irradiation and, hence, it might be universal for all iron-based superconductors.

In this work, we investigate by *ab initio* calculations the electronic structures of the (Lu;Y;Sc)$_2$Fe$_3$Si$_5$ superconductors and the non-superconducting isostructural Lu$_2$Ru$_3$Si$_5$ counterpart [29]. In our study, we are searching particularly for a possible relation between the FS topology and superconducting properties, in analogy to other two-gap superconductors as e.g. MgB$_2$ [33]. Finally, we discuss the differences occurring in the electronic structures between the rare-earth iron silicides and (oxy)pnictides or chalcogenides.

## 2. Computational methods

Electronic structure calculations of (Lu;Y;Sc)$_2$Fe$_3$Si$_5$ have been performed with the full-potential local-orbital (FPLO-9) method [34]. The Perdew-Wang form [35] of the local density approximation (LDA) of an exchange-correlation functional was employed in the scalar relativistic mode. The experimental x-ray diffraction values of lattice parameters of the unit cell (u.c.) having the *P4/mnc* symmetry for (Lu;Y;Sc)$_2$Fe$_3$Si$_5$ [2] and Lu$_2$Ru$_3$Si$_5$ [36] were used as the initial ones in further optimization of the u.c. volumes by minimizing the total energy - see Table 1. Here the u.c. contains four formula units (f.u.). The crystal structure is visualized in Fig. 1 where the same experimental atomic positions as obtained for Sc$_2$Fe$_3$Si$_5$ by the single-crystal x-ray refinement [37], have been assumed for all studied here iron-based systems. This assumption is justified by the fact that isoelectronic atoms, Lu, Y, and Sc, occupy equivalent positions in the u.c. and the experimental atomic positions of the considered Sc$_2$Fe$_3$Si$_5$ system [37] and e.g. Er$_2$Fe$_3$Si$_5$ [7] differ insignificantly, in spite of a considerable disparity in size between the Sc and Er atoms. The refined experimental atomic positions of Sc$_2$Fe$_3$Si$_5$ [37] were used as follows: Sc (Y;Lu): (0.0701, 0.2500, 0); Fe(1): (0, 1/2, 1/4); Fe(2): (0.3790, 0.3601, 0); Si(1): (0.1779, 0.6779, 1/4); Si(2)): (0, 0, 0.2528); Si(3): (0.1799, 0.4761, 0). For the reference Lu$_2$Ru$_3$Si$_5$ compound, the following atomic positions were taken from the work [36]: Lu: (0.075, 0.236, 0); Ru(1): (0, 1/2, 1/4); Ru(2): (0.371, 0.356, 0); Si(1): (0.185, 0.685, 1/4); Si(2): (0, 0, 0.242); Si(3): (0.191, 0.459, 0). The valence-basis sets have been selected by automatic procedure of the FPLO-9 code. The total energy values were converged with accuracy to ~1 meV for the 16x16x16 *k*-point mesh in the Brillouin zone (BZ), containing 621 points in its irreducible wedge.

## 3. Results and discussion

For the (Lu;Y;Sc)$_2$Fe$_3$Si$_5$ superconductors, the optimized by the LDA computations volumes of u.c., $V_{calc}$, amount to about 94-95 % of their experimental volumes, $V_{exp}$. It is worth underlining that the electronic structure results, especially the band energies and FS topology, obtained for $V_{calc}$ are almost the same as those yielded for $V_{exp}$. It turned out that any further changes are also negligible when simulating even much higher pressure than that used to achieve $V_{calc}$ starting from $V_{exp}$. Thus, we may assume that the large pressure effect on $T_C$, observed experimentally in the ternary iron silicides and reported in [38], might be connected rather with any electron-electron correlations than LDA electronic structure changes.

The total and partial (orbital *l* - resolved) densities of states (DOSs) of the studied systems are plotted in Fig. 2. As seen in this figure, their overall shape is similar for all three superconductors, differing mainly in the presence of a single narrow 4f-peak, existing only in the considered here Lu-based compounds. This peak is well localized, i.e. 4.5-5.0 eV below the Fermi level, $E_F$, and, therefore, it does not contribute to the DOS around $E_F$. In the reference system Lu$_2$Ru$_3$Si$_5$, the overall shape of the DOS, presented in the bottom part of Fig. 2, is much flatter than those in the three remaining superconductors. It is caused by the fact that the Ru 4d electron contribution is lying in a wider energy range below $E_F$ (down to -8 eV) than the Fe 3d electron contribution in (Lu;Y;Sc)$_2$Fe$_3$Si$_5$ (down to -5 eV). Moreover, in the latter three superconductors, the Fermi level is placed on the slope of a distinct peak consisting mainly of the Fe 3d states. This resembles the situation taking place in the common high-$T_C$ iron-based (oxy)pnictide supercondactors [39] but is does not concern the Ru 4d electron contributions in the case of Lu$_2$Ru$_3$Si$_5$. In all investigated here systems, the DOSs at $E_F$ have relatively low [~4-5 electrons /(eV*f.u.)] values, being typical of rather weak metallic systems. These densities around and well below $E_F$ are dominated either by the Fe 3d electrons, like it happens in the above mentioned high-$T_C$ superconductors [39], or by the Ru 4d electrons. In all cases, the contributions to DOS originating from both non-equivalent positions of iron/ruthenium atoms turned out to be almost equal to each other. In analogy, the contributions coming from all three different atomic sites of silicon atoms in the u.c. are quite comparable to one another. Only slight differences have been observed in electron population analysis given below.

Calculated electronic occupation numbers, $N_{calc}$, in the (Lu;Y;Sc)$_2$Fe$_3$Si$_5$ and Lu$_2$Ru$_3$Si$_5$ compounds compared to those for the free atoms, $N_{at}$, are collected in Table 2. As this table indicates, the electron populations for the Fe and Si atoms in all three superconductors are almost the same (within accuracy to ±0.1). There are only small differences (≤ 0.1) depending on their atomic position in the u.c. For the Fe 3d electrons, being the most responsible for SC, $N_{calc} \cong N_{at} + 1$, then, for the Lu 5d, Y 4d and Sc 3d states $N_{calc} \cong 1.5 N_{at}$ and, finally, for the Si 3d states (not present in free atoms) $N_{calc} \cong 0.5$. Moreover, the Si 3p states possess $N_{calc} > N_{at}$ whereas the valence p electrons of the Lu, Y, Sc and Fe atoms occur with $N_{calc} = 0.2$-$0.6$, although their $N_{at} = 0$. Oppositely, in all constituent atoms of the studied systems, their s electrons have $N_{calc} < N_{at}$.

Despite the fact that the occupation numbers $N_{at}$ of the Ru 4d and 5s electrons are different from the relative $N_{at}$ of the Fe 3d and 4s electrons, $N_{calc}$ of the above ruthenium states in Lu$_2$Ru$_3$Si$_5$, become similar to the corresponding iron states in the (Lu;Y;Sc)$_2$Fe$_3$Si$_5$ series (see Table 2). Also $N_{calc}$ of the other electron orbitals in Lu$_2$Ru$_3$Si$_5$ differ slightly (≤ 0.3) from the respective numbers $N_{calc}$ of the above iron-based superconductors.

For the (Lu;Y;Sc)2Fe$_3$Si$_5$ superconductors, we computed also the band weights, particularly weighted contributions of the Fe 3d states, having different orbital characters, to the energy bands. Since the Fe 3d weights are similar in all three systems, we have displayed them in Fig. 3 only for Lu$_2$Fe$_3$Si$_5$.

It is worth mentioning that the authors of the work [40] suggested that the unusual properties of the iron (oxy)pnictide superconductors originate from paring between the different Fe 3d orbitals of specific symmetry types. However, as seen in Fig. 3, the Fe 3d bands in Lu$_2$Fe$_3$Si$_5$ (and two remaining superconductors) reveal much higher hybridization of the Fe 3d orbitals with one another than that taking place in the (oxy)pnictides. It seems that the SC mechanism in these compounds might be still similar to that postulated earlier in the (oxy)pnictides but in this case the SC state can be considerably suppressed just due to the lack of a strong Fe 3d orbital separation. As is well known, also the parent compounds of high-$T_C$ superconducting (oxy)pnictides or chalcogenides exhibit comparably low $T_C$ values, e.g. FeSe ($T_C = 8$ K) [41]. It supports our assumption that although the quasi-2D electronic-structure properties of layered systems alone are not sufficient for occurring the unconventional SC, the quasi-2D features can be accountable for the sharp increase of $T_C$, observed under special conditions as doping (or external pressure) [1].

It is worth underlining that the calculated (LDA) Fermi surfaces of all three superconductors, (Lu;Y;Sc)$_2$Fe$_3$Si$_5$, are almost identical, in the scale of Fig. 4. Hence, we have drawn only the FS sheets for one representative, Lu$_2$Fe$_3$Si$_5$, together with the FS of non-superconducting Lu$_2$Ru$_3$Si$_5$. In this figure, the FS sheets of Lu$_2$Fe$_3$Si$_5$ originating from as many as four bands, are displayed only for three bands, denoted as I-III, since there are only negligibly small electron pockets in the higher band IV. The FS sheets confirm a fulfillment of conditions for multi-band SC in the (Lu;Y;Sc)$_2$Fe$_3$Si$_5$ family. This finding is also in good agreement with those reported in the previous work [21], where merely Lu$_2$Fe$_3$Si$_5$ was probed by the different from ours band-structure method (FLAPW) and also by employing various techniques of measurements. In analogy to MgB$_2$ [33], the postulated source of the multi-gap SC in Lu$_2$Fe$_3$Si$_5$ is first of all weak inter-band interactions between the multiple bands having different dimensionality [28]. In addition, the iron silicides exhibit moderately strong correlations between the Fe 3d electrons [29]. For the (Lu;Y;Sc)$_2$Fe$_3$Si$_5$ series, Fig. 4 points out that both the small holelike ellipsoidal pocket in the I FS sheet and large holelike cylindrical closed pocket in the II FS sheet, being located around the Γ point, have rather a 3D character. This is in contrary to the pronounced quasi-2D (along *c* axis) multi-band FS sheets in iron (oxy)pnictides or chalcogenides [39]. In the iron silicides, only the holelike I and II FS sheets, connected with the ZRA plane, and the electronlike III FS sheet exhibit some reduced dimensionality but along the *a* axis. This is better demonstrated by the FS section in the ΓXZ plane, as shown in Fig. 5.

The possible interaction just between these hole- and electronlike sheets having the different dimensionality in the studied here iron silicides suits well the discussed above picture of the multi-gap SC. Such an interaction enables also the extended s-wave (s± wave) superconductivity, for which the sign of the order parameter would be opposite for the hole- and electronlike FS sheets. This is consistent with recent experimental data [20,25,28-32], in analogy to the situation in the high-Tc (oxy)pnictides [42].

In the reference $Lu_2Ru_3Si_5$ compound, the FS has a similar overall character to that revealed in the $(Lu;Y;Sc)_2Fe_3Si_5$ superconductors. However, the holelike FS sheets contains the considerably smaller ellipsoidal pocket (centered at the Γ point) and there are small pillows instead of the larger I and II sheets (within the ZRA plane) existing in the above iron silicides. Also the holelike cylindrical II FS sheet is much flatter (along the ΓZ∥$c$ direction) than in the iron-based superconductors. What is more, the electronlike III FS sheet has a pronounced 3D closed shape, contrary to that found in the above iron silicides. Our results lead to the conclusion that the specific features of the electronic structures of the considered here $(Lu;Y;Sc)_2Fe_3Si_5$ should be crucial for arising their superconducting state.

## 4. Conclusions

Results of our band-structure calculations for the $(Lu;Y;Sc)_2Fe_3Si_5$ superconductors have shown almost identical their electronic structures, particularly the Fermi surfaces, but being distinctly different from those of the non-superconducting isostructural $Lu_2Ru_3Si_5$ counterpart. On the one hand, the specific properties of the Fermi surfaces in connection with moderately strong correlations of the Fe 3d electrons, suggested by the former experimental results, turned out to be crucial for multi-band superconductivity observed in these compounds, especially that these conditions are not fulfilled in the reference $Lu_2Ru_3Si_5$ compound. On the other hand, both the lack of a distinct orbital separation in the Fe 3d bands and a predominantly 3D character of the Fermi surfaces can be responsible for the considerably lower $T_C$ values observed in the $(Lu;Y;Sc)_2Fe_3Si_5$ superconductors compared to those of high-$T_C$ iron-based (oxy)pnictides or chalcogenides. Nevertheless, any unconventional (non-phononic) mechanism of their superconductivity may be similar to those already postulated for a majority of iron-based compounds. These conclusions are in accord with the findings of recent various experiments presented in [28-32].

**Acknowledgments**
The National Center for Science in Poland is acknowledged for financial support of Project No. N N202 239540. Calculations were carried out in Wroclaw Center for Networking and Supercomputing (Project No. 158). The Computing Center at the Institute of Low Temperature and Structure Research PAS in Wrocław is also acknowledged for the use of the supercomputers and technical support.

**References**

[1]  R.H. Liu, G. Wu, T. Wu, D.F. Fang, H. Chen, S.Y. Li, K. Liu, Y.L. Xie, X.F. Wang, R.L Yang, L. Ding, C. He, D.L. Feng, and X.H. Chen, Phys. Rev. Lett. **101**, 087001 (2008).
G. Wu, Y.L. Xie, H. Chen, M. Zhong, R.H. Liu, B.C. Shi, Q.J. Li, X.F. Wang, T. Wu, Y.J. Yan, J.J. Ying. and X.H. Chen, J. Phys.: Condens. Matter **21**, 142203 (2009).
[2]  H.F. Braun, Phys. Lett. **75A**, 386 (1980).
[3]  C.B. Vining, R.N. Shelton, H.F. Braun, and M. Pelizzone, Phys. Rev. B **27**, 2800 (1983).
[4]  A.R. Moodenbaugh, D.E. Cox. and H.F. Braun, Phys. Rev. B **25**, 4702 (1982).
[5]  D.R. Noakes, G.K. Shenoy, D. Niarchos, A.M. Umarji, and A.T. Aldred, Phys. Rev. B **27**, 4317 (1983).
[6]  C.B. Vining and R.N. Shelton, Phys. Rev. B **28**, 2732 (1983).
[7]  A.R. Moodenbaugh, D.E. Cox, C.B. Vining, and C.U. Segre, Phys. Rev. B **29**, 271 (1984).
[8]  A.R. Moodenbaugh, D.E. Cox, and C.B. Vining, Phys. Rev. B **32**, 3103 (1985).
[9]  Y. Singh, S. Ramakrishnan, Z. Hossain, and C. Geibel, Phys. Rev. B **66**, 014415 (2002).
[10] J.D. Casion, G.K. Shenoy, D. Niarchos, P.J. Viccaro, A.T. Aldred, and C.M Falco, J. Appl. Phys. **52**, 2180 (1981).
[11] H.F. Braun, C.U. Segre, F. Acker, M. Rosenberg, S. Dey, and P. Depp, J. Magn. Magn. Mater.

     **25**, 117 (1981).
- [12] H.H. Hamdeh, J.C. Ho, and H.D. Yang, Physica B **291**, 128 (2000).
- [13] S. Noguchi and K. Okuda, Physica B **194-196**, 1975 (1994).
- [14] C.U. Segre and H.F. Braun, Phys. Lett. **85A**, 372 (1981).
  C.B. Vining and R.N. Shelton, Solid State Commun. **54**, 53 (1985).
- [15] J.A. Gotaas, J.W. Lynn, R.N. Shelton, P. Klavins, and H.F. Braun, Phys. Rev. B **36**, 7277 (1987).
- [16] H. Schmidt, M. Müller, and H.F. Braun, Phys. Rev. B **53**, 12389 (1996).
- [17] Y. Singh and S. Ramakrishnan, J. Phys.: Condens. Matter **20**, 235243 (2008).
- [18] G.R. Stewart, G.P. Meisner, and C.U. Segre, J. Low. Temp. Phys. **59**, 237 (1985).
- [19] A. Mielke, W.W. Kim, J.J. Rieger, G. Fraunberger, E.W. Scheidt, and C.R. Stewart, Phys. Rev. B **50**, 16522 (1994).
- [20] T. Watanabe, H. Sasame, H. Okuyama, K. Takase, and Y. Takano, Phys. Rev. B **80**, 100502 (2009).
- [21] Y. Nakajima, T. Nakagawa, T. Tamegai, and H. Harima, Phys. Rev. Lett. **100**, 157001 (2008).
- [22] T. Tamegai, Y.T. Nakajima. Nakagawa, G.J. Li, and H. Harima, J. Phys.: Conf. Ser. **150**, 052264 (2009).
- [23] R.T. Gordon, M.D. Vannette, C. Martin, Y. Nakajima, T. Tamegai, and R. Prozorov, Phys. Rev. B **78**, 024514 (2008).
- [24] Y. Nakajima, G.J. Li, and T. Tamegai, Physica C **468**, 1138 (2008).
- [25] P.K. Biswas, G. Balakrishnan, D. McK. Paul, M.R. Lees, and A.D. Hillier, Phys. Rev. B **83**, 054517 (2011).
- [26] T. Tamegai, T. Nakagawa, and M. Tokunaga, Physica C **460-462**, 708 (2007).
- [27] T. Tamegai, Y. Nakajima, T. Nakagawa, G. Li, and H. Harima, Sci. Technol. Adv. Mater. **9**, 044206 (2008).
- [28] Y. Nakajima, H. Hidaka, T. Tamegai, T. Nishizaki, T. Sasaki, and N. Kobayashi, Physica C **469**, 921 (2009).
- [29] Y. Machida, S. Sakai, K. Izawa, H. Okuyama, and T. Watanabe, Phys. Rev. Lett. **106**, 107002 (2011).
- [30] H. Sasame, T. Masubuchi, K. Takase, Y. Takano, and T. Watanabe, J. Phys.: Conf. Ser. **150**, 052226 (2000).
- [31] H. Hidaka, Y. Nakajima, and T. Tamegai, Physica C **469**, 999 (2010).
  H. Hidaka, Y. Nakajima, and T. Tamegai, Physica C **470**, S619 (2010).
- [32] A.E. Karkin, M.R. Yangirov, N.Yu. Akshentsev, and B.N. Goshchitskii, Phys. Rev. B **84**, 054541 (2011).
- [33] J. Kortus, I.I. Mazin, K.D. Belashchenko, V.P. Antropov, and L.L. Boyer, Phys. Rev. Lett. **86**, 4656 (2001).
- [34] FPLO9.00-34, improved version of the FPLO code by K. Koepernik and H. Eschrig, Phys. Rev. B **59**, 1743 (1999); www.FPLO.de.
- [35] J.P. Perdew and Y. Wang, Phys. Rev. B **45**, 13244 (1992).
- [36] A.V. Morozkin and I.A. Sviridov, J. Alloys Compd. **296**, L4 (2000).
- [37] *O.I. Bodak, B.Y. Kotur, V.I. Yarovets, and E.I. Gladyshevskii, Sov. Phys. Crystallogr.* **22***, 217 (1977).*
- [38] *C.U. Segre and H.F. Braun, Physics of Solids Under High Pressure, edited by J.S. Schilling and R.N. Shelton (Amsterdam: North-Holland), p. 381 (1981).*
- [39] *D.J. Singh and M.-H. Du, Phys Rev. Lett.* **100***, 237003 (2008).*
  *S. Lebégue, Phys. Rev. B* **75***, 035110 (2007).*
  *C. Liu, G.D. Samolyuk, Y. Lee, N. Ni, T. Kondo, A.F. Santander-Syro, S.L. Bud'ko, J.L. McChesney, E. Rotenberg, T. Valla, A.V. Fedorov, P.C. Canfield, B.N. Harmon, and A. Kaminski, Phys. Rev. Lett.* **101***, 177005 (2008).*
- [40] S. Graser, T.A. Maier, P.J. Hirschfeld, and D.J. Scalapino, New J. Phys. **11**, 025016 (2009).
- [41] F.C. Hsu, J.Y. Luo, K.W. Yeh, T.K. Chen, T.W. Huang, P.M. Wu, Y.C. Lee, Y.L. Huang, Y.Y. Chu, D.C. Yan, and M.K. Wu, Proc. Natl. Acad. Sci. U.S.A. **105**, 14262 (2008).
- [42] I.I. Mazin, D.J. Singh, M.D. Johannes, and M.H. Du, Phys. Rev. Lett. **101**, 057003 (2008).
  K. Kuroki, S. Onari, R. Arita, H. Usui, Y. Tanaka, H. Kontani, and H. Aoki, Phys. Rev. Lett. **101**, 087004 (2008).

**Table 1.** Experimental and our calculated lattice parameters $a$ and $c$ (in nm) of $(Lu;Y;Sc)_2(Fe;Ru)_3Si_5$ compounds.

| compound | calculated, this paper | | experimental, Refs. [2] and [36] | |
| --- | --- | --- | --- | --- |
| | $a$ | $c$ | $a$ | $c$ |
| $Lu_2Fe_3Si_5$ | 1.0164 | 0.5284 | 1.0340 | 0.5375 |
| $Y_2Fe_3Si_5$ | 1.0221 | 0.5361 | 1.0430 | 0.5470 |
| $Sc_2Fe_3Si_5$ | 1.0000 | 0.5257 | 1.0225 | 0.5275 |
| $Lu_2Ru_3Si_5$ | - | - | 1.0611 | 0.5573 |

**Table 2.** Calculated occupation numbers, $N_{calc}$, (per given orbital at single atomic position) in the $(Lu;Y;Sc)_2(Fe;Ru)_3Si_5$ ternaries, compared to the corresponding numbers for free atoms, $N_{at}$ (with accuracy to ±0.1). Note that in each system, $N_{calc}$ of the same electron orbitals in Fe and Si atoms are varying slightly (±0.3) depending on their atomic positions in the u.c., from which given orbitals originate.

| $Lu_2Fe_3Si_5$ | Lu 5d 1.5 | Lu 6s 0.3 | Lu 6p 0.3 | Fe 3d 6.9 | Fe 4s 0.4-0.5 | Fe 4p 0.4-0.5 | Si 3s 1.3 | Si 3p 2.6-2.7 | Si 3d 0.4-0.5 |
| --- | --- | --- | --- | --- | --- | --- | --- | --- | --- |
| $Y_2Fe_3Si_5$ | Y 4d 1.5 | Y 5s 0.2 | Y 5p 0.2 | | | | | | |
| $Sc_2Fe_3Si_5$ | Sc 3d 1.6 | Sc 4s 0.4 | Sc 4p 0.6 | | | | | | |
| $Lu_2Ru_3Si_5$ | Lu 5d 1.4 | Lu 6s 0.3 | Lu 6p 0.3 | Ru 4d 7.0-7.2 | Ru 5s 0.4 | Ru 5p 0.3 | Si 3s 1.2-1.5 | Si 3p 2.6-2.8 | Si 3d 0.4 |
| $N_{at}$ | 1 | 2 | 0 | 6 (Fe) or 7 (Ru) | 2 (Fe) or 1 (Ru) | 0 | 2 | 2 | 0 |

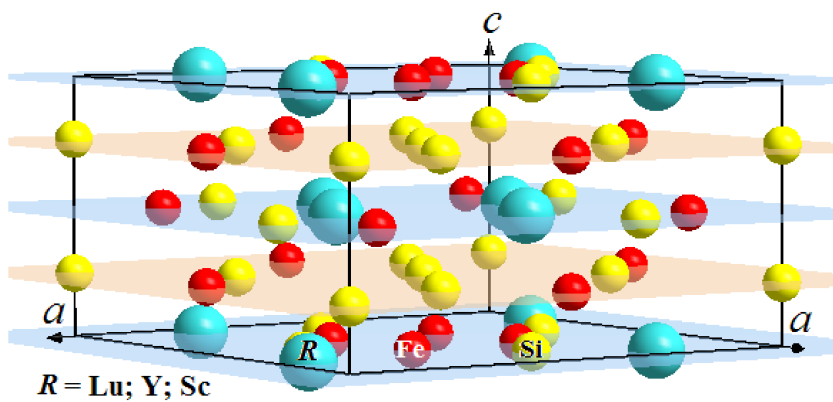

**Fig. 1** Tetragonal $P4/mnc$ crystal structure of the $Sc_2Fe_3Si_5$ type (no. 128).

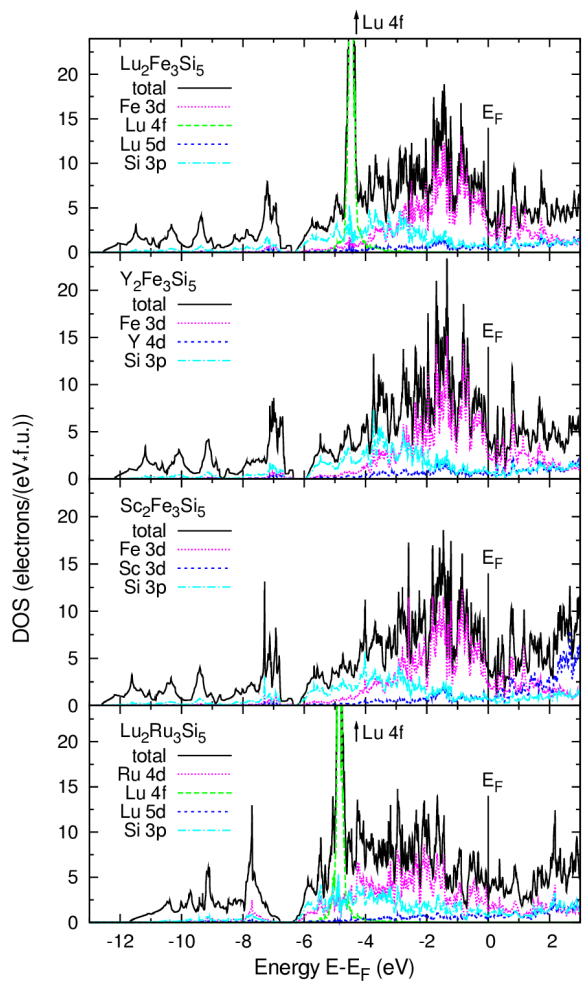

**Fig. 2** Computed total and partial (per electron orbitals of transition metal, 3d/4d/5d, and other, 3p/4p, atoms) DOSs, calculated (LDA) for the $R_2Fe_3Si_5$ ($R$ = Lu; Y; Sc) and $Lu_2Ru_3Si_5$ systems.

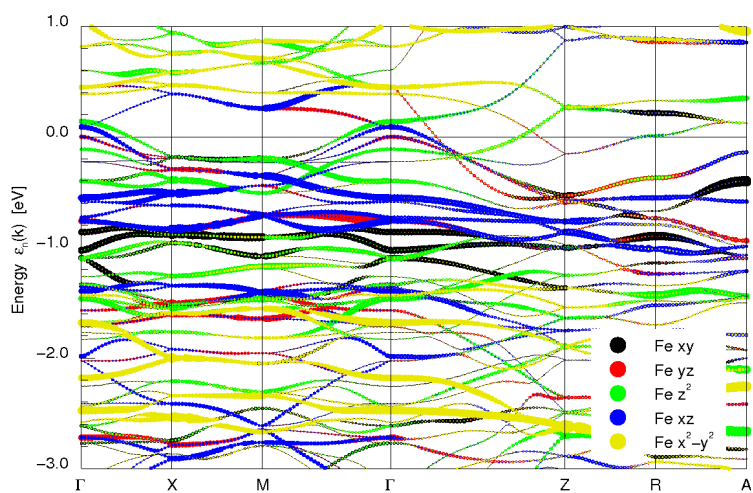

**Fig. 3** Computed (LDA) weighted bands in $Lu_2Fe_3Si_5$. The selected here predominant Fe 3d orbital characters are marked by circles of different colors. The circle sizes are proportional to given band weights of the orbitals.

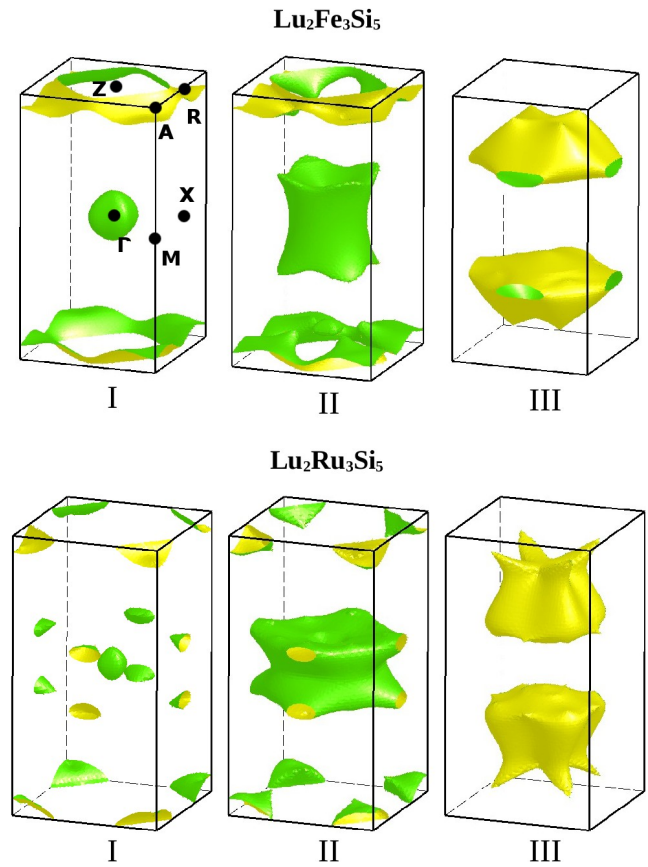

**Fig 4.** Calculated (LDA) for $Lu_2Fe_3Si_5$ and $Lu_2Ru_3Si_5$, main FS sheets, existing in three separate bands (denoted as I-III), drawn within the tetragonal BZ boundaries with marked high symmetry points. The FS sheets I-II and III have holelike and electronlike characters, respectively.

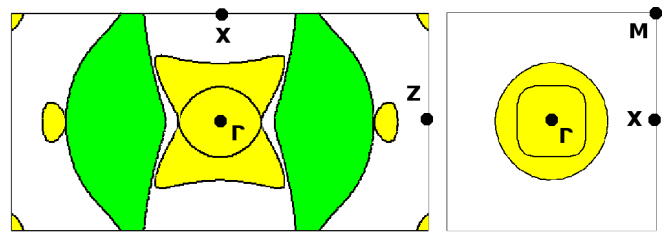

**Fig 5.** Sections through the FS sheets of $Lu_2Fe_3Si_5$, displayed in Fig. 4, drawn here together for I-III bands in the ΓXZ, (100) and ΓXM, (001), planes.